\documentclass[a4paper]{article}
\usepackage[dvipsnames]{xcolor}
\definecolor{mygray}{gray}{0.6}
\usepackage{INTERSPEECH2020}
\usepackage{cite}
\usepackage{bbding}
\title{MetricNet: Towards Improved Modeling For Non-Intrusive Speech Quality Assessment}
\name{Meng Yu, Chunlei Zhang, Yong Xu, Shixiong Zhang, Dong Yu}
\address{
  Tencent AI Lab, Bellevue, WA, USA}
\email{\{raymondmyu, cleizhang, lucayongxu, auszhang, dyu\}@tencent.com}

\begin{document}

\maketitle
\begin{abstract}
The objective speech quality assessment is usually conducted by comparing received speech signal with its clean reference, while human beings are capable of evaluating the speech quality without any reference, such as in the mean opinion score (MOS) tests. Non-intrusive speech quality assessment has attracted much attention recently due to the lack of access to clean reference signals for objective evaluations in real scenarios. In this paper, we propose a novel non-intrusive speech quality measurement model, MetricNet, which leverages label distribution learning and joint speech reconstruction learning to achieve significantly improved performance compared to the existing non-intrusive speech quality measurement models. We demonstrate that the proposed approach yields promisingly high correlation to the intrusive objective evaluation of speech quality on clean, noisy and processed speech data. 
\end{abstract}

\section{Introduction}
With the proliferation of hands-free devices and remote conferencing, speech-based human-machine interaction and voice communication become prevailing. The quality of speech signal is always degraded by environmental noises, room reverberation, hardware receivers, digital signal processing, and networking transmission. Speech quality assessment is in high demand in those applications. For example, the environment sensing function tells a device whether it sits in a noisy acoustic environment so that environment-aware modeling or environment dependent feedback will be applied. The voice communication service providers seek to estimate the perceived audio signal quality for monitoring the audio service to their customers \cite{Falk06}. 

A few metrics have been widely used to estimate the speech quality, such as signal-to-noise ratio (SNR), signal-to-distortion ratio (SDR), signal-to-interference ratio (SIR) and signal-to-artifact ratio (SAR) \cite{Vincent06}. These metrics offer the objective measures of target speech over a variety of other non-target factors. However, they may not correlate well with human perception of speech quality. The subjective listening test is considered the best way for speech quality evaluation. 
The ITU-T standardized the perceptual evaluation of speech quality by introducing the 
MOS \cite{ITU-T98}. Human listeners are asked to judge the quality of speech utterances in a scale from 1 to 5. A MOS is obtained by averaging all participants’ scores over a particular condition. However, such subjective measurements are not always feasible due to the time and labor costs \cite{Avila16} and may vary for the same clip from one evaluator to another \cite{Reddy20}. Alternatively, 
objective measurement such as perceptual evaluation of speech quality (PESQ), was introduced by ITU-T Recommendation P.862 \cite{ITU-T01}, followed by the perceptual objective
listening quality assessment (POLQA) from ITU-T Recommendation P.863 as an update to PESQ particularly for super-wideband \cite{ITU-T11}. PESQ and POLQA predictions principally model MOS. 

The objective assessment of speech quality can be divided into two ways, intrusive and non-intrusive \cite{ITU-T04}, depending on whether the clean reference exists. Those mentioned above such as SNR related, PESQ and POLQA are all intrusive measurements. Although intrusive methods are considered more accurate as they provide a higher correlation with subjective evaluations, the requirement of the clean reference channel limits their applications to a lot of real scenarios. In contrast, non-intrusive measurements use only the corrupted speech signal for
quality assessment. It performs like human beings, forming a mapping from the perceived speech utterance to a quality score. 

\begin{figure*}[!t]
        \centering
        \vspace{-2cm}
        \includegraphics[width=\linewidth]{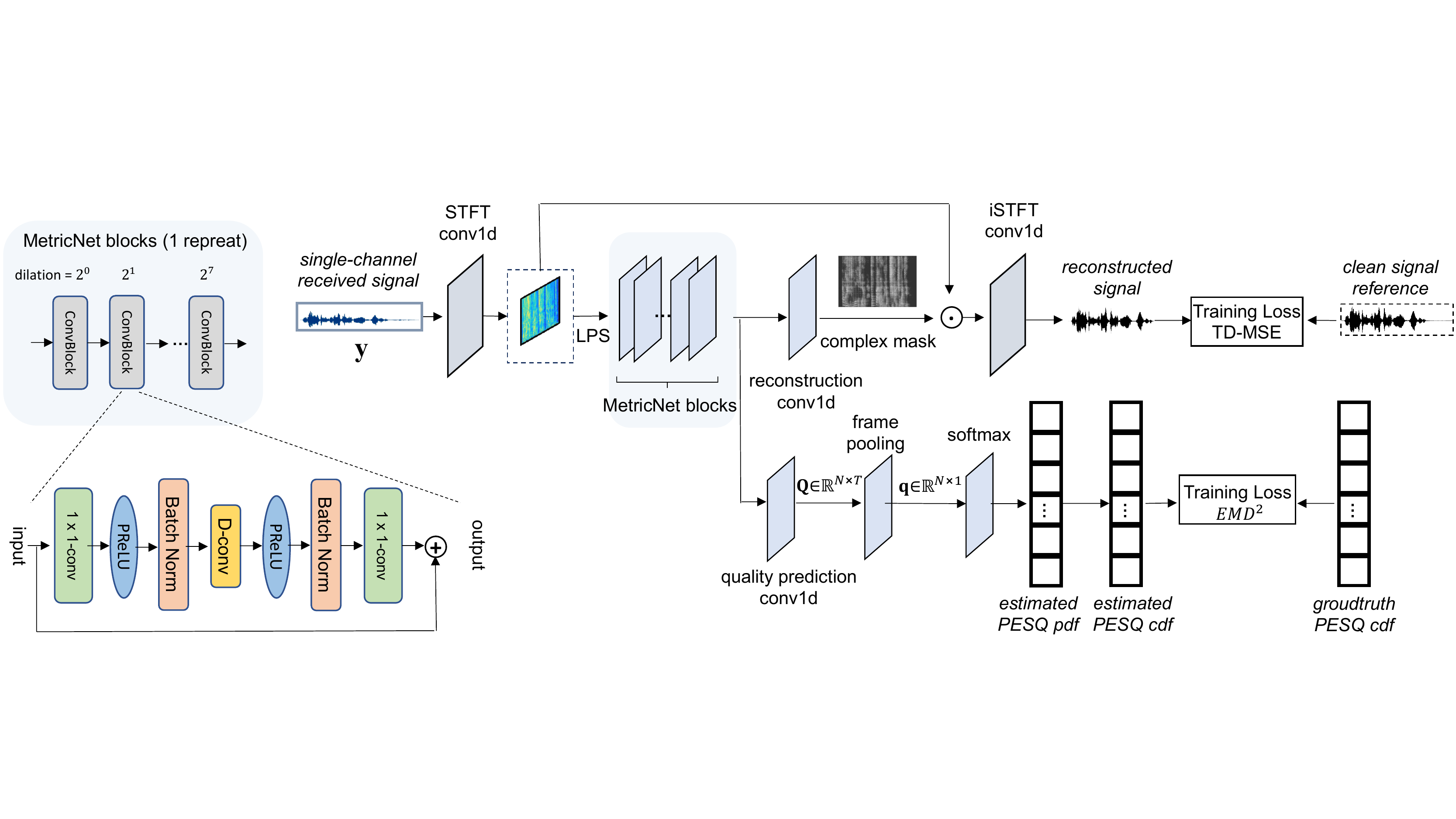}
        \vspace{-2.5cm}
        \caption{The overall MetricNet model architecture and the convolutional structure of MetricNet blocks. Each block (ConvBlock) consists of a 1$\times$1-conv layer, a depth-wise separable convolution layer ($D-conv$), with PReLU activation function and normalization added between each two convolution layers. The residual connection is added in each block.}
        \label{fig:model}
        \vspace{-0.5cm}
\end{figure*}

ITU-T Recommendation P.563 \cite{ITU-T04} is a non-intrusive technique which directly operates on the degraded audio signal. However, it was developed for narrow-band applications only. A handful of machine learning based techniques have been used for conducting non-intrusive metric learning in the past decade \cite{Falk06, Grancharov06, Ding07, Narwaria12, Dubey13, Sharma14, Hakami17}. Those complex handcrafted features used in the metric learning are not jointly optimized with the models. Meanwhile, such metric models are not differentiable, which limits their capability to work with other systems in a joint optimization manner. Inspired by the great success of deep learning, the deep neural networks have been developed to address the non-intrusive speech evaluation problem recently \cite{Soni16, Spille18, Fu18, Andersen18, Lo19, Gamper19, Avila19, Dong19, Dong20, Reddy20}. In \cite{Fu18}, an end-to-end and non-intrusive speech
quality evaluation model, termed Quality-Net, was proposed based on bidirectional long short-term memory (BLSTM). The Quality-Net was applied to predict the PESQ scores for both noisy and enhanced speech signals. Several neural network architectures were investigated in MOSNet \cite{Lo19} where MOS was used as the training target for predicting the human MOS ratings, including convolutional
neural network (CNN), BLSTM, and CNN-BLSTM, as these architectures have shown their capability to model human perception. In \cite{Gamper19}, Gamper et al. performed a regression tree analysis to predict the average perceptual ratings and generate the training labels. In \cite{Avila19} a few input features to the neural networks for score prediction were compared. \cite{Dong20} presented a unified model that leverages different aspects of speech assessment. A multi-stage self-teaching model was presented in \cite{Reddy20} for improved leaning in the presence of noisy labels by crowd sourcing. 

Despite the recent attention on applying deep learning to solve this problem, few research aims to improve the modeling and training criterion for this particular task. The common training criterion for speech quality estimation is the
mean square error (MSE) or mean absolute error (MAE) between the predicted value and the ground-truth. Such methods usually cannot achieve satisfactory results because some outliers may cause a large error term which affects training stability. In this paper, we transform the regression based PESQ estimation which minimizes the MSE to a multi-class single-label classification problem. By viewing the multi-class prediction as the label distribution learning \cite{Geng16}, the real value PESQ score is converted to a discrete PESQ distribution. Although softmax cross-entropy loss is by far the most popular loss
function for such training regime, this loss function does not take into account inter-class relationships which can be very informative \cite{Gao18, Hou16}. Therefore, we refer to the earth mover’s distance (EMD) \cite{Rubner00}, as known as the Wasserstein distance \cite{Bogachev12}, to compute the minimal cost required to transform the predicted distribution to the ground truth label distribution. Furthermore, as the essential principle of speech quality measurement is to find the level of perceptual degradation, we equip the quality score prediction network with a speech reconstruction sub-network, which estimates the clean speech signal from its noisy or degraded version. We show that solving such inverse problem helps to identify speech degradation level. The speech reconstruction layers are then jointly optimized with the quality prediction network. 

Note that although PESQ is selected as the training target, the proposed method serves as a preliminary work and can be extended to work with other large-scaled human labeled subjective metrics. 
The rest of the paper is organized as follows. In Section \ref{sec:metricnet}, we first recap two related methods, and then present MetricNet. We describe our experimental setups and evaluate the effectiveness of the proposed system in Section \ref{exp}. We conclude this work in Section \ref{con}.

\section{MetricNet}\label{sec:metricnet}
\subsection{Recap of Related Methods}\label{ov}
Quality-Net \cite{Fu18} learns frame-level as well as utterance-level qualities. The ultimate utterance-level quality score is estimated by combining the frame-wise scores through a global average. The overall loss function consists of two parts with respect to utterance-level and frame-level estimations, respectively. 
Since the inconsistency of utterance-level and frame-level scores becomes insignificant when the quality of the input utterance is high, it assigns a higher/lower weight to the frame-level term when the quality of the input utterance is higher/lower. 
With the extra frame-wise regularization, it is shown to achieve promisingly better results than utterance-level only scheme. 

It was also noticed in \cite{Dong19} that by minimizing the MSE the regression task may lead to prediction outliers and cause the model to overfit. By doing a very coarse quantization on real value PESQ labels, a classification loss was added to the MSE based regression loss of PESQ for punishing samples with large estimation errors. 

\subsection{MetricNet Model}\label{sub:metricNet}
The overall architecture of the proposed MetricNet is shown in Fig.\ref{fig:model}. Our previous work in \cite{bahmaninezhad2019comprehensive,Gu20} implemented the multi-channel speech enhancement network by a dilated CNN similar to conv-TasNet \cite{luo2019convtasnet} but through a short-time Fourier transform (STFT) for signal encoding. Such network structure supports a long reception field to capture more contextual information and is thus adopted in the MetricNet. 
The MetricNet starts from an encoder that maps the single-channel input waveform to complex spectrogram by a STFT 1-D convolution layer. The STFT window size is 32 ms and the hop size is 16 ms. Based on the complex spectrogram $\mathbf{Y}$, the logarithm power spectrum (LPS) is extracted as $\text{LPS} = \log(|\mathbf{Y}|^2) \in \mathbb{R}^{F\times T}$, where $T$ and $F$ are the total number of frames and frequency bands of the complex spectrogram, respectively. The $\text{LPS}$ feature is then passed to the MetricNet blocks, which consist of stacked dilated convolutional layers with exponentially growing dilation factors \cite{luo2019convtasnet}. 
The detailed design of these MetricNet blocks follows \cite{luo2019convtasnet}, as illustrated in Fig.\ref{fig:model}. The number of channels in $1\times1\textit{-conv}$ layer is set as 256. For the $D-conv$ layer, the kernel size is 3 with 512 channels. Batch normalization is applied. Every 8 convolutional blocks are packed as a repeat, with exponentially increased dilation factors $2^0, 2^1, ..., 2^7$. Four repeats in total are used in those blocks. 

The output of MetricNet blocks goes to two branches. In the signal reconstruction branch, two conv1d layers of 256 channels map the output of MetricNet blocks to real and imaginary parts of a complex mask, respectively. The predicted complex mask is multiplied with the input complex spectrogram $\mathbf{Y}$. An inverse STFT (iSTFT) 1-D convolution layer converts the reconstructed speech complex spectrogram to the waveform $\hat{\mathbf{x}}$. 

In the quality prediction branch, a conv1d layer of $N$ channels maps the output of MetricNet blocks to $\mathbf{Q}\in \mathbb{R}^{N\times T}$, where $N$ sets the number of partitions in a certain range of the training target. For PESQ, the range is from -0.5 to 4.5 according to P.862 \cite{ITU-T01}. A equal step size $\Delta l = 5/N$ is used to quantize PESQ into a vector $[-0.5: \Delta l: 4.5]$. Therefore, the $n$-th class corresponds to the PESQ value falling to the interval $(l_{n-1} : l_n] :=(-0.5+(n-1)*\Delta l : -0.5+n*\Delta l]$. Frame pooling computes an average over time frames by $\mathbf{q} = \frac{1}{T}\sum_{t=1}^T\mathbf{Q}_t$, where $\mathbf{q} \in \mathbb{R}^{N\times 1}$ and $\mathbf{Q}_t$ denotes the $t$-th column of $\mathbf{Q}$, i.e. the $t$-th frame's output in an utterance. The softmax is applied to compute the estimated posterior distribution across all $N$ classes as
\begin{equation}
    \hat{p}_n  = \frac{e^{q_n}}{\sum_{k=1}^{N}e^{q_k}}.
\end{equation}
The goal of MetricNet is to maximize the similarity between the predicted distribution $\hat{\mathbf{p}}$ and the ground-truth probability density function (PDF) $\mathbf{p}$ of PESQ over all $N$ classes in the training stage. During the inference stage, the predicted distribution $\hat{\mathbf{p}}$ is reversed to a real valued PESQ by either picking the middle point of the interval corresponding to the maximum value in $\hat{p}_n, n = 1, 2, \cdots, N$ as shown in (\ref{eq:pesq_peak}), or an expected value over the predicted distribution by (\ref{eq:pesq_ws})
\vspace{-0.3cm}
\begin{align}
&\hat{S}_{M}  =-0.5+(\arg\max_n \hat{\mathbf{p}}-1)*\Delta l + \frac{\Delta l}{2}  \label{eq:pesq_peak}\\
&\hat{S}_{E}  = \sum_{n=1}^N \hat{p}_n * [-0.5+(n-1)*\Delta l + \frac{\Delta l}{2}] \label{eq:pesq_ws}
\end{align}

%



\subsection{Training Criterion}\label{subsec:mlen}
Although the scale-invariant signal-to-distortion ratio (SI-SDR) has been used as the objective function to optimize the enhancement network in \cite{luo2019convtasnet, bahmaninezhad2019comprehensive, Gu20}, showing superior to frequency domain MSE loss, a potential drawback of SI-SDR is pointed in \cite{Roux19} that it does not consider scaling as an error. As a result, SI-SDR will lead to an uncontrollable scaling in the output of MetricNet blocks, which is not desirable for the quality score prediction. We thus use a time-domain MSE (TD-MSE) defined in (\ref{eq:mse}) as a relevant measure

\vspace{-0.1cm}
\begin{equation}
    \text{TD-MSE}(\hat{\mathbf{x}}, \mathbf{x}):=
\left\|\mathbf{x} - \hat{\mathbf{x}}\right\|_{2}^{2},
\label{eq:mse}
\end{equation}
where $\hat{\mathbf{x}}$ and $\mathbf{x}$ are the estimated and clean target speech waveforms, respectively. The zero-mean normalization is applied to both $\hat{\mathbf{x}}$ and $\mathbf{x}$.

\begin{figure}[t]
  \includegraphics[width=150mm, height=80mm]{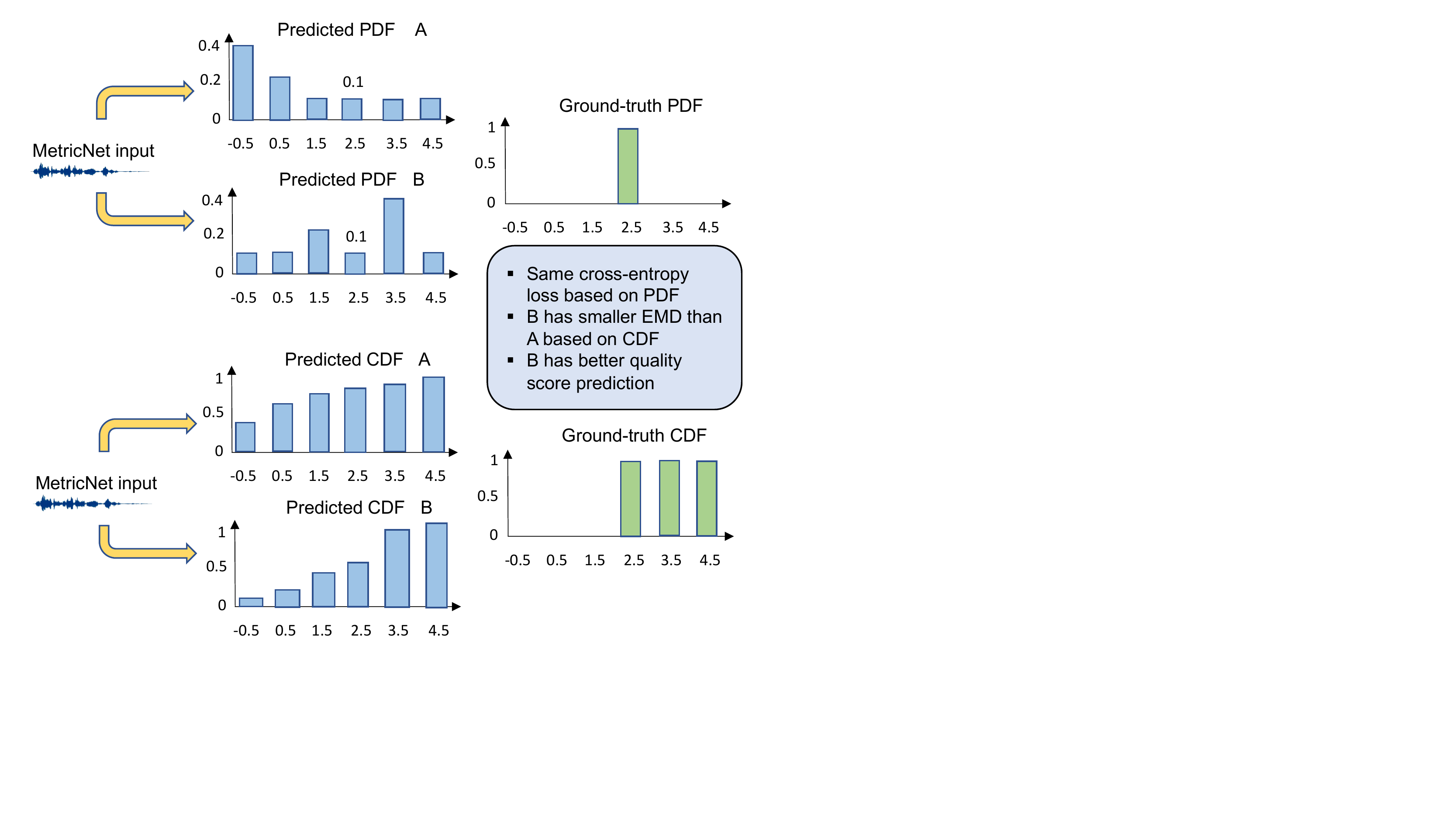}
  \vspace{-2.1cm}
  \caption{An example that two predicted distributions have the equal cross-entropy loss while B is preferable to A, clearly seen from their EMD loss}
\vspace{-0.5cm}
  \label{loss}
\end{figure}

Quality estimation is a very fine-grained task, e.g., even human listeners hardly sense the change of speech degradation level from for example 2.5 to 2.4. For reducing the risk of overfitting and large estimation outliers, we transform the regression task for a scalar estimation to a classification task for predicting the quality score in one of $N$ classes $(l_{n-1}, l_n], n = 1, 2, \cdots, N$. As the classes are ordered, the inter-class relationships are very informative. However, the cross-entropy loss only focuses on the predicted ground-truth class while ignores the relationships between classes. An example is illustrated in Fig.\ref{loss}. For quality score estimation, a nearby prediction around the ground-truth is preferable to a far away prediction. We aim to minimize the distance from the predicted distribution to the ground-truth labeled distribution. Such label distribution learning has been studied in \cite{Geng16} and achieved the state-of-the-art performance on age estimation \cite{Gao18}. The EMD metric finds the minimal cost required to transform one distribution to another \cite{Levina01}. EMD has a closed-form solution if the classes are naturally in a sorted order \cite{Levina01}. A simple solution is given by the squared error between the cumulative distribution functions(CDF) of the prediction and the label distributions in \cite{Hou16}. 
\vspace{-0.3cm}
\begin{equation}
    \textbf{EMD}^2(\hat{\mathbf{p}}, \mathbf{p}) = \sum_{n=1}^N \left ( \hat{P}_n  -  P_n  \right )^2,
\end{equation}
where $P_n$ is the $n$-th element of the cumulative density function of $\mathbf{p}$. The label distribution typically follows a one-hot vector where the element 1 is at $\nu$-th position if the ground-truth quality score falls into the interval $(l_{\nu-1}, l_\nu]$. The probability
density function of normal distribution is also a natural
choice to generate the label distribution. For simplicity, we use the following discrete distribution as a soft label distribution \cite{Subramanian21} compared to the one-hot one:
\begin{align}
p_n = \begin{cases}
0.4 & n= \nu,\\
0.2 & n=\nu \pm 1,\\
0.1 & n=\nu \pm 2,\\
0 & elsewhere,
\end{cases}
\label{soft}
\end{align}
where $\nu$ is the index corresponding to the ground-truth quality score class. Note that we add extra classes at the PESQ range boundaries for using this soft label distribution.  

The MetricNet simultaneously reconstructs the clean target speech and predicts the quality of the received speech signal. The loss function thereby becomes
\vspace{-0.1cm}
\begin{equation}
\vspace{-0.1cm}
\mathcal{L}=
\text{TD-MSE}(\hat{\mathbf{x}}, \mathbf{x}) + \text{EMD}^2(\hat{\mathbf{p}} ,\mathbf{p})
\label{eq:totalLoss}
\end{equation}

\begin{table*}[t]
\centering
\caption{\label{eval1} {\it Evaluation on clean, reverberant, noisy and processed data }}
\vspace{-0.2cm}
\begin{tabular}{l|cccc|ccc|ccc}\toprule

 \midrule  Method  & $N$ & label dist. & rankLoss & Joint Recon. & MSE &  LCC &  SRCC & MSE &  LCC &  SRCC  \\
  \midrule    Classification aided \cite{Dong19}& n/a & n/a & n/a &  n/a& 0.126& 0.925 & 0.940 &n/a & n/a& n/a\\
  Quality-Net utt. only \cite{Fu18} & n/a  & n/a & n/a & n/a & 0.164 & 0.921 & 0.932  &n/a &n/a &n/a \\
   Quality-Net frame reg.\cite{Fu18} & n/a & n/a & n/a &  n/a& 0.109 & 0.933 &0.951 &n/a & n/a& n/a\\
    \midrule  &  & & & & \multicolumn{3}{c|}{Expectation score $\hat{S}_E$} & \multicolumn{3}{c}{Max. likelihood $\hat{S}_M$}\\
   \midrule (a) MetricNet & 100 & one-hot & \texttimes & \texttimes &  0.112 & 0.933 & 0.947 & 0.123& 0.925& 0.939 \\
   (b) MetricNet & 500 & one-hot & \texttimes&  \texttimes  & 0.095 & 0.942 & 0.953 &0.103 & 0.936& 0.949\\
   (c) MetricNet & 100 & soft & \texttimes&  \texttimes  & 0.086 & 0.947 & 0.957 & 0.091&0.944 &0.955\\
   (d) MetricNet & 100 & one-hot & \checkmark & \texttimes  &  0.105 & 0.936 & 0.952 & 0.111&0.932 &0.949 \\
   (e) MetricNet & 100 & soft & \checkmark & \texttimes  &  0.086 & 0.948 & 0.957 & 0.091 & 0.944 & 0.955 \\
   (f) MetricNet & 100 & one-hot &\texttimes & \checkmark  &  \textbf{0.079} & \textbf{0.952} & \textbf{0.962} & \textbf{0.081} & \textbf{0.951} & \textbf{0.961}\\
   (g) MetricNet & 100 & soft & \texttimes & \checkmark  &  0.082 & 0.951 & 0.960& 0.086 & 0.948 & 0.960\\
   (h) MetricNet & 100 &one-hot& \checkmark & \checkmark &  0.081& 0.951 & 0.962& 0.085& 0.949&0.961 \\
   (i) MetricNet & 100 & soft &\checkmark & \checkmark &  0.084& 0.950 & 0.960&0.086 & 0.949& 0.960\\
\bottomrule
\end{tabular}
\end{table*}

\begin{table*}[t]
\centering
\caption{\label{eval2} {\it Evaluation on augmented processed data by one-hot label distribution and expectation based score $\hat{S}_E$ }}
\vspace{-0.2cm}
\begin{tabular}{l|cccccc}\toprule
 \midrule  Method  & $N$ & rankLoss & Joint Recon. & MSE &  LCC &  SRCC \\
\midrule Classification aided \cite{Dong19} & n/a & n/a & n/a & 0.105 & 0.869  & 0.899   \\
   Quality-Net utt. only \cite{Fu18} & n/a & n/a & n/a & 0.111 & 0.892 & 0.897 \\
   Quality-Net frame reg. \cite{Fu18} & n/a & n/a &  n/a& 0.079 & 0.914 &0.915 \\
   (a) MetricNet & 100 & \texttimes & \texttimes &  0.077 & 0.891 & 0.907 \\
   (b) MetricNet & 500 & \texttimes&  \texttimes  & 0.069 & 0.910 & 0.915\\
   (c) MetricNet & 100 & \checkmark & \texttimes  &  0.072 & 0.900 & 0.914 \\
   (d) MetricNet & 100 & \texttimes & \checkmark  &  \textbf{0.060} & \textbf{0.918} & \textbf{0.931} \\
   (e) MetricNet & 100 & \checkmark & \checkmark &  0.063 & 0.916 & 0.927 \\
\bottomrule
\end{tabular}
\vspace{-0.4cm}
\end{table*}

\section{Experiments}\label{exp}
\subsection{Experitmental Setup}\label{subsec:kws}
We simulate a single-channel reverberant and noisy dataset by AISHELL-2 corpus \cite{Du2018}.
The room simulator based on the image method \cite{Allen79} generates 10K rooms with random room characteristics, speaker and microphone locations. The corresponding room sizes (length$\times$width$\times$height) range from $3m\times3m\times2.5m$ to $8m\times10m\times6m$. The reverberation time $T_{60}$s range from 0 to 600 ms, with an average $T_{60}$ of 300 ms. The simulated room impulse responses (RIRs) are randomly selected for creating reverberant waveforms.
The environmental point and diffuse noise sources are mixed with the simulated utterances with SNR randomly sampled from -12dB to 30dB. We generate 90K utterances from 1800 speakers, 7.5K utterances from 150 speakers and 2K utterances from 40 speakers for training, validation and testing, respectively. The speakers in training, validation and testing are not overlapped. Besides the clean, reverberant and noisy utterances, two neural networks for denoising and denoising \& dereverberation, respectively, are used to process the generated dataset. For training, validation and testing, we sample the claimed number of utterances over the original generated set, denoised set and denoised \& dereverberated set at a ratio 50\%, 25\%, and 25\%, respectively.  
To evaluate the performance of MetricNet, MSE, linear correlation coefficient (LCC), and
Spearman’s rank correlation coefficient (SRCC) \cite{Myers03} are computed between the predicted and ground-truth PESQ scores. For a fair comparison, the baseline models in \cite{Fu18, Dong19} are re-implemented by using the same ConvBlocks as we use in MetricNet. Other parts are kept the same as those in the original papers. All models are trained with the same data.

\subsection{Results and Discussion}
Table \ref{eval1} shows the evaluation results on the generated clean, reverberant, noisy and processed 
speech data. By comparing the ways that quality scores are computed, expectation on the predicted distribution outperforms maximum likelihood value in all setups. [a vs b] Finer quantization on PESQ from $N=100$ to $500$, i.e. minimum class size from $0.05$ to $0.01$, reduces the MSE from 0.112 to 0,095 and improves the correlation to ground-truth PESQ as well. [a vs c, d vs e] The soft label distribution defined in (\ref{soft}) leads to better performance for MetricNet without the speech reconstruction branch. The soft label distribution helps the model to learn better inter-class relationships. SRCC measures the rank correlation between predicted scores and ground-truth scores. For instance, if an utterance has higher PESQ than another, it is expected to have higher MetricNet score as well. We therefore integrate such meaningful rank metric into our objective function. [a vs d, c vs e] MetricNet without speech reconstruction learning gets subtle improvement by incorporating the rank loss. [e] Up to now, the presented MetricNet with only the label distribution learning reduces the MSE by 21\% compared to the best baseline Quality-Net with frame regularization. [a vs f, c vs g] The joint training with speech reconstruction obtains clear performance gains for both label distribution settings. [f vs g, h vs i] For MetricNet with speech reconstruction learning, the smoother label distribution does not benefit the system. The learning process for recovering the clean speech signal drives the model to predict sharper distributions. [f vs h, g vs i] Similarly, the rank loss does not bring extra gains in this setup neither. Overall, the MetricNet with both label distribution learning and speech reconstruction learning reduces the MSE by 28\% and as well as significantly improves the output correlation to the target quality scores. 

Furthermore, we generate perturbations on the processed testing data by randomly boosting 30\% and attenuating 50\% time-frequency bins of spectrograms. Such perturbation on spectrogram bins generates more unseen artifacts. The perturbed data are used for the evaluation in Table \ref{eval2}. Different setups come to the same conclusions as those observed in Table \ref{eval1}. MetricNet outperforms the best baseline by reducing MSE from $0.079$ to $0.060$.

\section{Conclusions}\label{con}
In this paper we designed a novel model, named MetricNet, for improved learning of non-intrusive speech quality measurement. Compared to the existing deep learning based approaches, our contribution is two-fold, building speech signal reconstruction into the end-to-end MetricNet for better resolving speech quality degradation level and secondly incorporating the earth mover’s distance for label distribution learning as our training criterion. Experimental results showed that the proposed MetricNet yields a significantly higher correlation to the training target PESQ and outperforms the baseline Quality-Net and a classification aided method. As a preliminary work, the presented model and training criterion can be readily used to work with other learning labels for speech assessment tasks. 

\newpage
\bibliographystyle{IEEEtran}
\bibliography{mybib}

\begin{thebibliography}{10}
\providecommand{\url}[1]{#1}
\csname url@samestyle\endcsname
\providecommand{\newblock}{\relax}
\providecommand{\bibinfo}[2]{#2}
\providecommand{\BIBentrySTDinterwordspacing}{\spaceskip=0pt\relax}
\providecommand{\BIBentryALTinterwordstretchfactor}{4}
\providecommand{\BIBentryALTinterwordspacing}{\spaceskip=\fontdimen2\font plus
\BIBentryALTinterwordstretchfactor\fontdimen3\font minus
  \fontdimen4\font\relax}
\providecommand{\BIBforeignlanguage}[2]{{%
\expandafter\ifx\csname l@#1\endcsname\relax
\typeout{** WARNING: IEEEtran.bst: No hyphenation pattern has been}%
\typeout{** loaded for the language `#1'. Using the pattern for}%
\typeout{** the default language instead.}%
\else
\language=\csname l@#1\endcsname
\fi
#2}}
\providecommand{\BIBdecl}{\relax}
\BIBdecl

\bibitem{Falk06}
W.~H. Falk and W.-Y. Chan, ``Single-ended speech quality measurement using
  machine learning methods,'' \emph{IEEE Transactions on Audio, Speech, and
  Language Processing}, vol.~14, no.~6, pp. 1935--1947, 2006.

\bibitem{Vincent06}
E.~Vincent, R.~Gribonval, and C.~Fevotte, ``Performance measurement in blind
  audio source separation,'' \emph{IEEE Transactions on Audio, Speech, and
  Language Processing}, vol.~14, no.~4, 2006.

\bibitem{ITU-T98}
ITU-T, ``Recommendation p.800: Methods for subjective determination of
  transmission quality,'' \emph{ITU-T Recommendation P.800}, 1998.

\bibitem{Avila16}
A.~R. Avila, B.~Cauchi, S.~Goetze, S.~Doclo, and T.~Falk, ``Performance
  comparison of intrusive and non-intrusive instrumental quality measures for
  enhanced speech,'' in \emph{IEEE International Workshop on Acoustic Signal
  Enhancement (IWAENC)}, 2016.

\bibitem{Reddy20}
C.~K. Reddy, V.~Gopal, and R.~Cutler, ``Dnsmos: A non-intrusive perceptual
  objective speech quality metric to evaluate noise suppressors,'' \emph{arXiv
  e-prints}, pp. arXiv--2010, 2020.

\bibitem{ITU-T01}
ITU-T, ``Recommendation p.862: Perceptual evaluation of speech quality (pesq),
  an objective method for endto-end speech quality assessment of narrowband
  telephone networks and speech codecs,'' 2001.

\bibitem{ITU-T11}
------, ``Recommendation p.863: Perceptual objective listening quality
  assessment: An advanced objective perceptual method for end-to-end listening
  speech quality evaluation of fixed, mobile, and ip-based networks and speech
  codecs covering narrowband, wideband, and super-wideband signals,'' 2011.

\bibitem{ITU-T04}
------, ``Recommendation p.563: Single-ended method for objective speech
  quality assessment in narrowband telephony applications,'' \emph{ITU-T
  Recommendation P.563}, 2004.

\bibitem{Grancharov06}
V.~Grancharov, D.~Y. Zhao, J.~Lindblom, and W.~B. Kleijn, ``Low-complexity,
  nonintrusive speech quality assessment,'' \emph{IEEE Transactions on Audio,
  Speech, and Language Processing}, vol.~14, pp. 1948--1956, 2006.

\bibitem{Ding07}
L.~Ding, Z.~Lin, A.~Radwan, M.~S. El-Hennawey, and R.~A. Goubran,
  ``Non-intrusive single-ended speech quality assessment in voip,''
  \emph{Speech communication}, vol.~49, pp. 477--489, 2007.

\bibitem{Narwaria12}
M.~Narwaria, W.~Lin, I.~V. McLoughlin, S.~Emmanuel, and L.-T. Chia,
  ``Nonintrusive quality assessment of noise suppressed speech with
  mel-filtered energies and support vector regression,'' \emph{IEEE
  Transactions on Audio, Speech, and Language Processing}, vol.~20, pp.
  1217--1232, 2012.

\bibitem{Dubey13}
R.~K. Dubey and A.~Kumar, ``Non-intrusive speech quality assessment using
  several combinations of auditory features,'' \emph{International Journal of
  Speech Technology}, vol.~16, pp. 89--101, 2013.

\bibitem{Sharma14}
D.~Sharma, L.~Meredith, J.~Lainez, D.~Barreda, and P.~A. Naylor, ``A
  non-intrusive pesq measure,'' in \emph{IEEE Global Conference on Signal and
  Information Processing (GlobalSIP)}, 2014.

\bibitem{Hakami17}
M.~Hakami and W.~B. Kleijn, ``Machine learning based nonintrusive quality
  estimation with an augmented feature set,'' in \emph{IEEE International
  Conference on Acoustics, Speech and Signal Processing (ICASSP)}, 2017, pp.
  5105--5109.

\bibitem{Soni16}
M.~H. Soni and H.~A. Patil, ``Novel deep autoencoder features for non-intrusive
  speech quality assessment,'' in \emph{24th European Signal Processing
  Conference (EUSIPCO)}, 2016, pp. 2315--2319.

\bibitem{Spille18}
C.~Spille, S.~Ewert, B.~Kollmeier, and B.~Meyer, ``Predicting speech
  intelligibility with deep neural networks,'' \emph{Computer Speech and
  Language}, vol.~48, pp. 51--66, 2018.

\bibitem{Fu18}
S.-W. Fu, Y.~Tsao, H.-T. Hwang, and H.-M. Wang, ``Quality-net: An end-to-end
  non-intrusive speech quality assessment model based on blstm,'' in
  \emph{Proc. Interspeech}, 2018.

\bibitem{Andersen18}
A.~H. Andersen, J.~M. Haan, Z.~Tan, and J.~Jensen, ``Non-intrusive speech
  intelligibility prediction using convolutional neural networks,'' \emph{IEEE
  Transactions on Audio, Speech, and Language Processing}, vol.~26, pp.
  1925--1939, 2018.

\bibitem{Lo19}
C.-C. Lo, S.-W. Fu, W.-C. Huang, X.~Wang, J.~Yamagishi, Y.~Tsao, and H.-M.
  Wang, ``Mosnet: Deep learning-based objective assessment for voice
  conversion,'' in \emph{Proc. Interspeech}, 2019.

\bibitem{Gamper19}
H.~Gamper, C.~Reddy, R.~Cutler, I.~Tashev, and J.~Gehrke, ``Intrusive and
  non-intrusive perceptual speech quality assessment using a convolutional
  neural network,'' in \emph{IEEE Workshop on Applications of Signal Processing
  to Audio and Acoustics}, 2019.

\bibitem{Avila19}
A.~Avila, H.~Gamper, C.~Reddy, R.~Cutler, I.~Tashev, and J.~Gehrke,
  ``Non-intrusive speech quality assessment using neural networks,'' in
  \emph{IEEE International Conference on Acoustics, Speech and Signal
  Processing (ICASSP)}, 2019.

\bibitem{Dong19}
X.~Dong and D.~S. Williamson, ``A classification-aided framework for
  non-intrusive speech quality assessment,'' in \emph{IEEE Workshop on
  Applications of Signal Processing to Audio and Acoustics}, 2019.

\bibitem{Dong20}
------, ``An attention enhanced multi-task model for objective speech
  assessment in real-world environments,'' in \emph{IEEE International
  Conference on Acoustics, Speech and Signal Processing (ICASSP)}, 2020.

\bibitem{Geng16}
X.~Geng, ``Label distribution learning,'' \emph{IEEE Transactions on Knowledge
  and Data Engineering}, vol.~28, pp. 1734--1748, 2016.

\bibitem{Gao18}
B.~Gao, H.~Zhou, J.~Wu, and X.~Geng, ``Age estimation using expectation of
  label distribution learning,'' in \emph{The 27th International Joint
  Conference on Artificial Intelligence (IJCAI)}, 2018.

\bibitem{Hou16}
L.~Hou, C.-P. Yu, and D.~Samaras, ``Squared earth mover's distance-based loss
  for training deep neural networks,'' \emph{arXiv:1611.05916}, 2016.

\bibitem{Rubner00}
Y.~Rubner, C.~Tomasi, and L.~J. Guibas, ``The earth mover’s distance as a
  metric for image retrieval,'' in \emph{International Journal of Computer
  Vision}, 2000, pp. 99--121.

\bibitem{Bogachev12}
V.~I. Bogachev and A.~V. Kolesnikov, ``The mongekantorovich problem:
  achievements, connections, and perspectives,'' \emph{Russian Mathematical
  Surveys}, vol.~67, 2012.

\bibitem{bahmaninezhad2019comprehensive}
F.~Bahmaninezhad, J.~Wu, R.~Gu, S.-X. Zhang, Y.~Xu, M.~Yu, and D.~Yu, ``A
  comprehensive study of speech separation: spectrogram vs waveform
  separation,'' \emph{Proc. Interspeech}, 2019.

\bibitem{Gu20}
R.~Gu, S.-X. Zhang, Y.~Xu, L.~Chen, Y.~Zou, , and D.~Yu, ``Multi-modal
  multi-channel target speech separation,'' \emph{IEEE Journal of Selcted
  Topics in Signal Processing,}, 2020.

\bibitem{luo2019convtasnet}
Y.~{Luo} and N.~{Mesgarani}, ``Conv-tasnet: Surpassing ideal time–frequency
  magnitude masking for speech separation,'' \emph{IEEE/ACM Transactions on
  Audio, Speech, and Language Processing}, vol.~27, no.~8, pp. 1256--1266, Aug
  2019.

\bibitem{Roux19}
J.~L. Roux, S.~Wisdom, H.~Erdogan, and J.~R. Hershey, ``Sdr - half-baked or
  well done?'' in \emph{IEEE International Conference on Acoustics, Speech and
  Signal Processing (ICASSP)}, 2019.

\bibitem{Levina01}
E.~Levina and P.~Bickel, ``The earth mover’s distance is the mallows
  distance: some insights from statistics,'' in \emph{IEEE International
  Conference on Computer Vision}, 2001.

\bibitem{Subramanian21}
A.~Subramanian, C.~Weng, S.~Watanabe, M.~Yu, and D.~Yu, ``Deep learning based
  multi-source localization with source splitting and its effectiveness in
  multi-talker speech recognition,'' \emph{arXiv:2102.07955}, 2021.

\bibitem{Du2018}
J.~Du, X.~Na, X.~Liu, and H.~Bu, ``Aishell-2: Transforming mandarin asr
  research into industrial scale,'' \emph{arXiv:1808.10583}, 2018.

\bibitem{Allen79}
J.~B. Allen and D.~A. Berkley, ``Image method for efficiently simulation
  room-small acoustic,'' \emph{Journal of the Acoustical Society of America},
  vol.~65, no.~4, pp. 943--950, 1979.

\bibitem{Myers03}
J.~L. Myers and A.~D. Well, ``Research design and statistical analysis (2nd
  ed.),'' \emph{ISBN 978-0-8058-4037-7.}, 2003.

\end{thebibliography}

\end{document}